# Observation of whistler wave instability driven by temperature anisotropy of energetic electrons on EXL-50 spherical torus

Mingyuan Wang[1,2,3], Yuejiang Shi[2,3*], Jiaqi Dong[2,3*], Xinliang Gao[4], Quanming Lu[4], Ziqi Wang[2,3], Wei Chen[5], Adi Liu[4], Ge Zhang[4], Yumin Wang[2,3], Shikui Cheng[2,3,5], Mingsheng Tan[6], Songjian Li[2,3], Shaodong Song[2,3], Tiantian Sun[2,3], Bing Liu[2,3], Xianli Huang[2,3], Yingying Li[2,3], Xianming Song[2,3], Baoshan Yuan[2,3], Y-K Martin Peng[2,3], and ENN team.

1 School of Mathematics and Physics, Anqing Normal University, Anqing, 246133, People's Republic of China

2 Hebei Key Laboratory of Compact Fusion, Langfang 065001, China

3 ENN Science and Technology Development Co., Ltd., Langfang 065001, China

4. University of Science and Technology of China, Anhui Hefei 230026, China

5 Southwestern Institute of Physics, Chengdu 610041, China

6 Institute of Energy, Hefei Comprehensive National Science Center, Hefei 230031, China

*E-mail of corresponding author: yjshi@ipp.ac.cn,  jiaqi@swip.ac.cn

**Abstract**

Electromagnetic modes in the frequency range of 30-120MHz were observed in electron cyclotron wave (ECW) steady state plasmas on the ENN XuanLong-50 (EXL-50) spherical torus. These modes were found to have multiple bands of frequencies proportional to the Alfvén velocity. This indicates that the observed mode frequencies satisfy the dispersion relation of whistler waves. In addition, suppression of the whistler waves by the synergistic effect of Lower Hybrid Wave (LHW) and ECW was also observed. This suggests that the whistler waves were driven by temperature anisotropy of energetic electrons. These are the first such observations (not runaway discharge) made in magnetically confined toroidal plasmas and may have important implications for studying wave-particle interactions, RF wave current driver, and runaway electron control in future fusion devices.

**Introduction**

Whistler waves, one of fundamental plasma waves, have been widely observed in space and laboratory, which are important for controlling electron dynamics. It is widely accepted that the wave-particle interaction of whistler waves is the dominant mechanism for energizing electrons in the radiation belt [1-4]. In laboratory experiments, whistler waves have been excited using methods such as electron beams, whistler wave antennas, and runaway electrons to investigate the potential physical mechanisms of whistler wave instabilities [3, 5-14]. Low frequency whistler waves ($\omega < 0.1\omega_{ce}$, $\omega_{ce}$ is the electron cyclotron frequency) driven by runaway electrons of parallel temperature ($T_\parallel$) higher than perpendicular temperature ($T_\perp$) has been observed in tokamak [12, 13]. However, low-frequency whistler wave instabilities induced by temperature anisotropy of energetic electrons with $T_\parallel$ lower than $T_\perp$ have not been explicitly observed in tokamaks. In this study, we present the first experimental observation of whistler wave instability driven by the temperature anisotropy of energetic electrons in electron cyclotron wave (ECW) steady-state plasmas. These observations, conducted on a steady-state operating experimental platform, provide valuable insight into the physical mechanisms of wave-particle interaction and have important implications for understanding the ionosphere of the Earth, Van Allen betas, and tokamak reactors.

Firstly, this paper reports on high frequency mode excitation experiments conducted on the EXL-50 spherical torus and presents the main characteristics of the mode in $\Lambda > 0$, ($\Lambda = T_\perp/T_\parallel - 1$) steady-state electron cyclotron wave (ECW) plasmas. The experimental results showed that the mode frequencies were found to be greater than the ion cyclotron frequency ($\omega_{ci}$), but much smaller than $\omega_{ce}$. Additionally, it was observed that the mode frequencies were proportional to the Alfvén velocity ($V_A$) and that the mode featured multiple frequency bands. These findings suggest that this high frequency mode is whistler wave [9, 12].

Secondly, mechanism of the whistler wave instability is explored. The results of plasma density scanning experiment indicate that as the plasma density increases, the whistler wave is gradually suppressed and the loss of energetic electrons is also mitigated. The whistler waves were observed to be suppressed when the plasma density was greater than $2 \times 10^{18} m^{-2}$. As the whistler waves were observed in ECW heating plasma, it is speculated that they are driven by the electron temperature anisotropy. To further verify this, a Lower-hybrid wave (LHW) was used to increase the parallel electron temperature in the ECW plasma. The result was of significant whistler wave suppression. On the other hand, disruptions associated with whistler waves were observed, which implies that these waves may trigger magnetic reconnection [15-17]. These studies contribute to the understanding of whistler wave physics in tokamaks and provide valuable insights for the research on RF current drive and control of runaway instability.

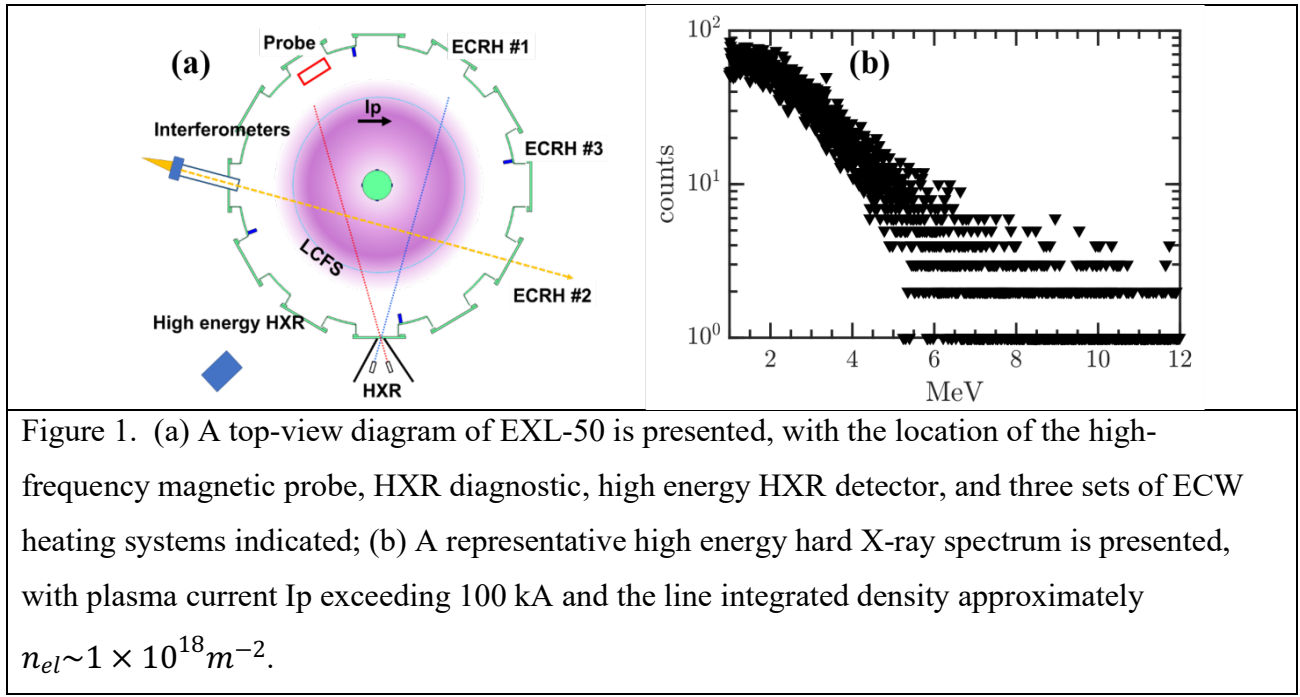

Figure 1. (a) A top-view diagram of EXL-50 is presented, with the location of the high-frequency magnetic probe, HXR diagnostic, high energy HXR detector, and three sets of ECW heating systems indicated; (b) A representative high energy hard X-ray spectrum is presented, with plasma current Ip exceeding 100 kA and the line integrated density approximately $n_{el} \sim 1 \times 10^{18} m^{-2}$.

**EXL-50 experiment setup**

The whistler wave experiments were performed on EXL-50 spherical tours using a ECW at 28GHz with a power of approximately 200 kW and LHW at 2.45GHz with a power of approximately 60 kW in low electron density plasmas (the line integrated density $n_{el} \sim 1 - 2 \times 10^{18} m^{-2}$) [18]. EXL-50 is a medium-sized spherical torus without a central solenoid (CS), featuring a major 0.58 m, a minor radius of approximately 0.41 m, a toroidal magnetic field ($B_T$) of approximately 0.5 T at R ~ 0.48 m, and an aspect ratio of A > 1.45. Unless otherwise specified, only the ECW was used to heat and maintain the plasma. Under these conditions, $T_\perp \gg T_\parallel$ is achieved [19] and a large number of energetic electrons are generated by the ECW to start-up [20] and maintain plasma currents [21]. The whistler wave fluctuations, associated with sufficient temperature anisotropy (especially energetic electrons), were excited [2, 7, 22, 23] and measured using a high-frequency magnetic probe on the low field-side of the mid-plane [24] (as illustrated in Figure 1(a)). The energetic electron loss to the wall was measured using CdZnTe detectors [25] and a LaBr3(Ce) scintillation detector. The LaBr3(Ce) scintillation detector was placed approximately 20 m away from the torus center, and a typical hard-X ray spectrum is shown in Figure 1(b).

## Experimental results and analysis

The time evolution of magnetic fluctuation power spectra (a) and associated parameters of typical ECW discharges (b) are depicted in Figure 2. Multiple coherent modes with frequencies ranging from 70 to 110 MHz (shot # 15033) and 30-120 MHz (shot # 14375) are observed. The frequency of the modes is observed to decrease gradually as the plasma density increases and the magnetic field strength decreases.

The parametric behavior of the whistler wave frequency can be understood through the dispersion relation in cold plasma, as outlined in equation [9, 26]:

$$\omega = kV_A\sqrt{1 + k_\parallel^2 c^2/\omega_{pi}^2}, \qquad (1)$$

where $\omega$ is the wave frequency, $\bm{V_A} = B_T/\sqrt{4\pi n_e m_i}$ is the Alfvén velocity, $n_e$ is the electron density and $m_i$ is the ion mass, $\bm{\omega_{pi}}$ is the ion diamagnetic frequency, $\bm{k}$ and $\bm{k_\parallel}$ are the total and component parallel to the magnetic field of the wave vector, respectively. For the lowest order ($k_\parallel^2 c^2/\omega_{pi}^2 \sim 0$), Eq.(1) indicates a linear scaling of the frequency with Alfvén velocity ($\omega \propto V_A$) and an inverse square root scaling with the plasma density ($\omega \propto n_e^{-1/2}$).

These trends are confirmed with the results of density and magnetic field scanning experiments in Figures 2 (c) and (d), where mode frequency is clearly shown to be inversely proportional to the square root of the plasma density (Figure 2 c) and linearly proportional to the Alfvén velocity (Figure 2 d). It suggests that the coherent modes are whistler waves. It is worth pointing out that the frequencies of the observed whistler waves are much lower than the ECW frequency ($\omega \sim 0.01\omega_{ce}$).

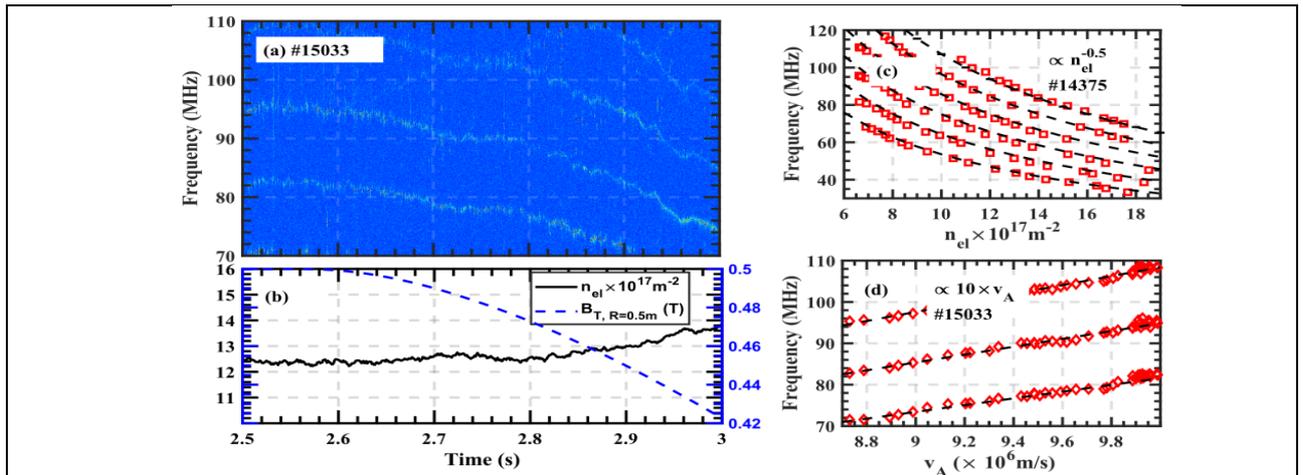

Figure 2. Time evolution of (a) magnetic fluctuation power spectra; (b) toroidal magnetic field (near the magnetic axis), and line integrated density. Variation of mode frequency spectrum with line integrated density at constant toroidal magnetic field (c) and Alfvén velocity (d).

Experiments and theories show that a stochastic wave can accelerate electrons to MeV energy at an input ECW power of a few 100 kW [27, 28]. Photons of 6 MeV energy are measured under steady-state plasma conditions by high-energy hard X-ray diagnostics with ECW injected power >200 kW in the experiment, which indicates that energetic electrons can excite the instabilities by wave-particle interaction. The observed whistler wave instability may be driven when energetic electrons resonate with the wave [12]. The resonance condition is,

$$\omega - \boldsymbol{k}_\parallel \boldsymbol{v}_\parallel - k_\perp v_d - l\omega_{ce}/\gamma = 0.$$

Here, $\boldsymbol{v}_\parallel$ is the parallel velocity of the electrons, $k_\perp$ is the perpendicular component of the wave-vector, $v_d$ is the orbital drift velocity, and $\gamma$ is the relativistic factor. In the case of EXL-50, the observed wave frequency is much lower than the electron cyclotron frequency; therefore, $l = -1$ or $l = 0$ resonances are relevant.

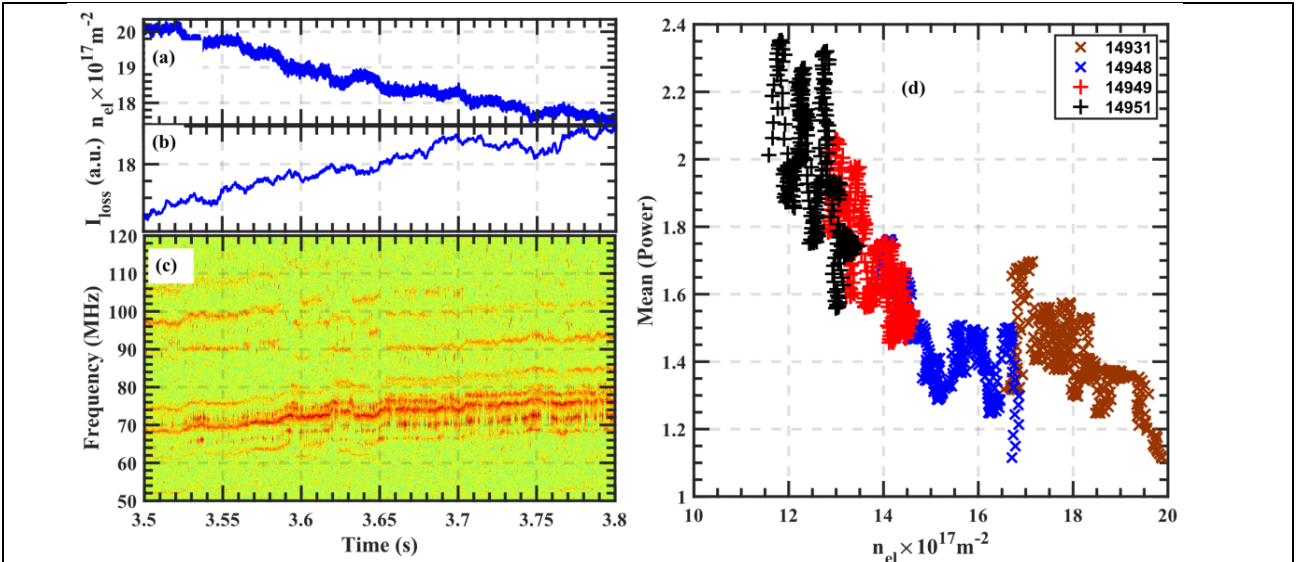

Figure 3. Time evolution of (a) the line-integrated electron density, (b) distant hard X-ray signal, (c) whistler wave spectra, and (d) intensity of whistler wave fluctuations versus electron density, there is a threshold value of the electron density, triggering the whistler wave.

The excitation of whistle wave ($\omega_{ci} \ll \omega \ll \omega_{ce}$) instability is triggered by temperature anisotropy when $\Lambda > \frac{\omega}{\omega_{ce}}$ [29]. This instability induces enhancement of fluctuating fields, which in turn lead to wave particle scattering, thereby reducing the temperature anisotropy. The minimum anisotropy required to trigger these processes corresponds to the onset of strong wave-particle interactions, which generally represents a constraint on the anisotropy [2, 4, 30-32]. The electron temperature anisotropy $T_\perp/T_\parallel$ is bounded by the threshold condition of the whistler instability:

$$\Lambda \propto \gamma/\beta_\parallel .$$

Here, $\beta_{e\parallel} = 8\pi n_e T_{e\parallel}/B_T^2$ is the parallel beta of the energetic electron, and $\gamma$ is fitting parameter that is related to the wave growth rate. Although the observed frequency of the whistler wave is significantly lower than the electron cyclotron frequency, the observed relationships of the whistler wave intensity being inversely proportional to density and positively proportional to temperature anisotropy are still consistent with the scenario where the whistler wave is driven by temperature anisotropy.

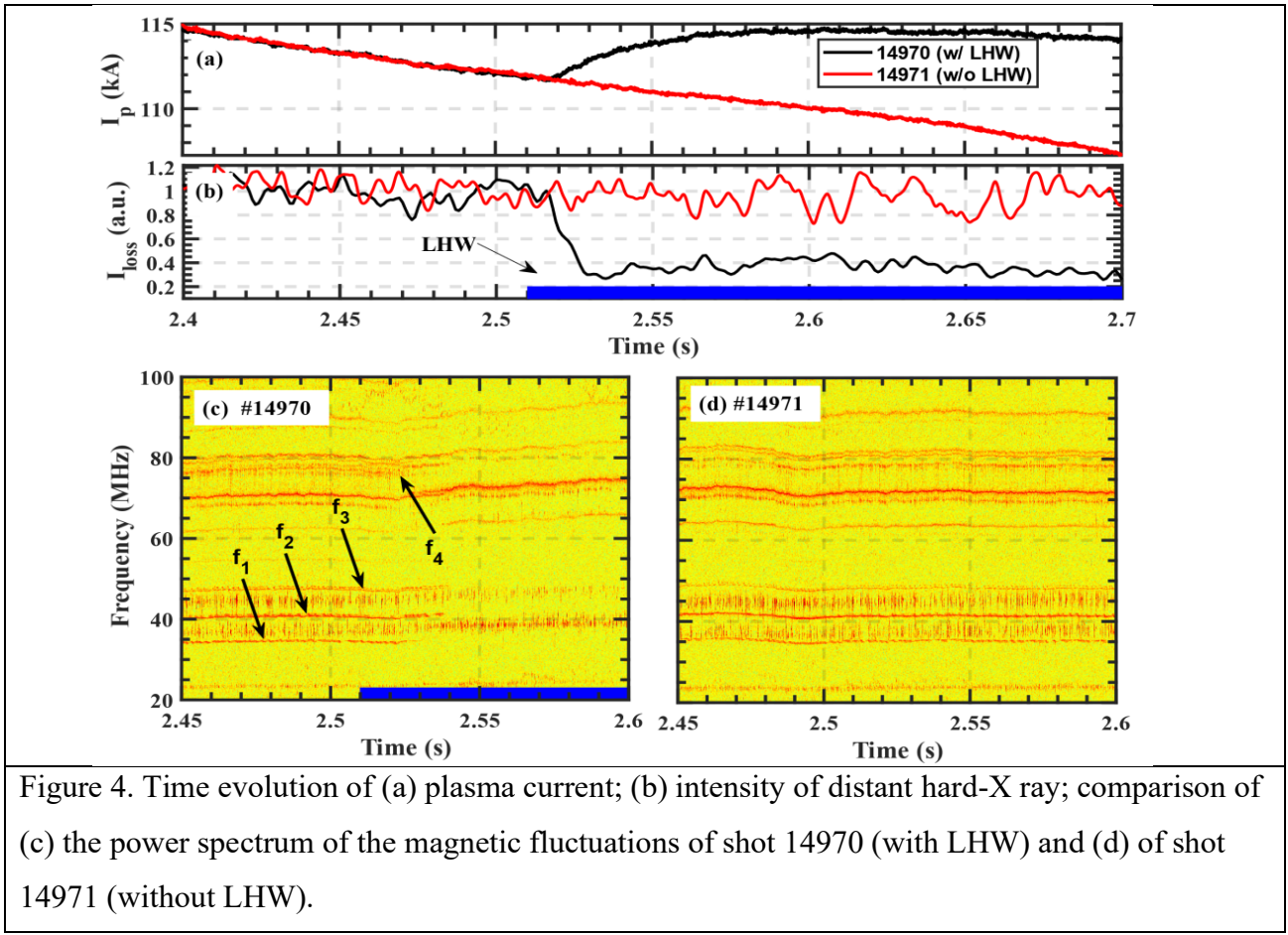

Figure 4. Time evolution of (a) plasma current; (b) intensity of distant hard-X ray; comparison of (c) the power spectrum of the magnetic fluctuations of shot 14970 (with LHW) and (d) of shot 14971 (without LHW).

Here, we explore the correlation between the energy of saturated wave magnetic field and the bulk electron density. Figure 3 (c) presents time evolutions of the intensity and frequency of the whistler wave corresponding to the evolutions of plasma density shown in Fig. 3(a) and the intensity of hard X-ray signal induced by last electrons shown in Fig. 3(b). As the plasma density decreases, the intensity and frequency of the whistler wave simultaneously increase. Figure 3(d) displays density threshold ($2 \times 10^{18} m^{-2}$) for the instability. Finally, it is worth pointing out that the enhancement of whistle wave intensity is correlated with an increase of hard X-ray radiation induced by energetic electron loss (Fig.3(c)), indicating that the whistle wave scatters energetic electrons into the loss phase space [33, 34].

Currently, direct measurements of the parallel and perpendicular temperatures of energetic electrons are not available on EXL-50. Therefore, LHW was employed as a qualitative approach to verify the temperature anisotropy threshold for the whistler instability. In shot 14970, LHW was turned on at 2.52s, and shot 14971 served as a reference discharge without LHW. As shown in Figure 4(a), the plasma current increased after the LHW activation in shot 14970, while the plasma current decreased continuously without the LHW in shot 14971. The increase in the plasma current indicates an increase in the parallel energy of the electrons driven by the LHW. Additionally, the distant HXR intensity is significantly reduced after the LHW activation (Figure 4 (b)), suggesting a reduction in energetic electron loss, which may be attributed to instability mitigation. The power spectrum of high-frequency fluctuation before and after LHW activation is compared (Figure 4 (c)). The bands with frequency rise and fall ($f_1$, f2, f3, f4) are completely suppressed, when LHW is activated. Whereas the intensities of the bands with frequency rise and fall remain almost unchanged for shot 14971 (see Figure 4 (d)). This represents the first observation of whistler waves driven by electron temperature anisotropy on an ST.

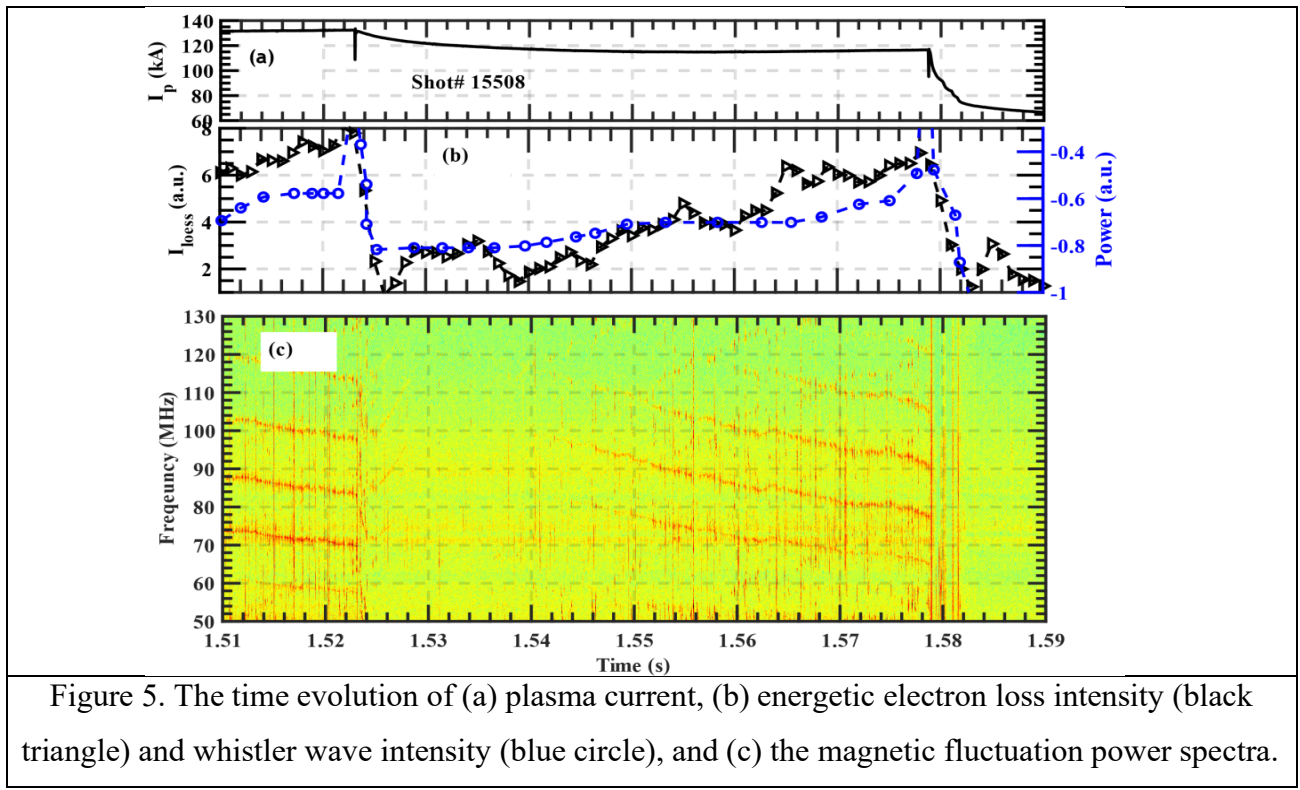

Figure 5. The time evolution of (a) plasma current, (b) energetic electron loss intensity (black triangle) and whistler wave intensity (blue circle), and (c) the magnetic fluctuation power spectra.

Theoretical and experimental studies have demonstrated that scattering of whistler instability wave on electrons in homogeneous plasma can maintain the distribution of electrons near instability thresholds [2]. However, in toroidal plasma, the whistler wave may alter equilibrium by modifying the macroscopic properties (current and pressure) of the plasma. Although the ELMO Bumpy Torus exhibits a relatively low level of plasma current, disruptions have been observed [14] when the high-frequency hot electron instability is strongly excited.

As shown in Figure 5, plasma current disruptions were observed when the intensity of whistler wave reached a certain level. The radiation intensity induced by loss of energetic electrons increases with the whistler wave intensity increase, and current disruptions occur when the whistler wave intensity reaches a certain threshold,. These findings suggest that ECW may excite such waves and potentially limit the efficiency of current drive. Given that ECW non-inductive current drive is crucial for the advancement of spherical torus fusion devices, this is an issue that requires significant attention in ST.

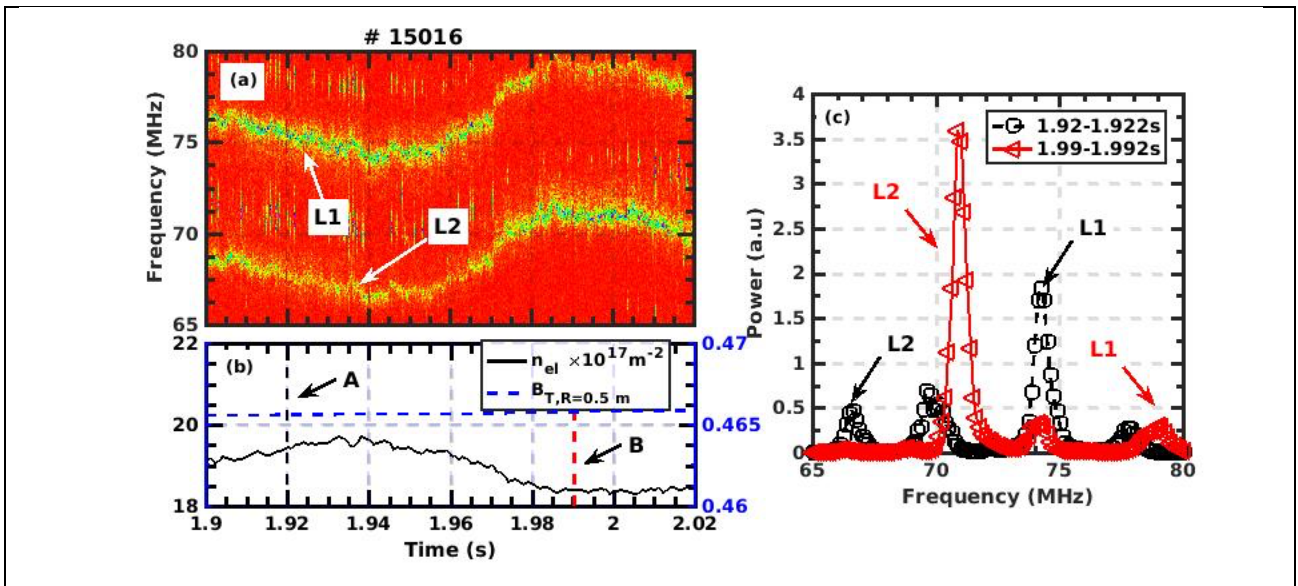

Figure6. The time evolution of (a) high frequency magnetic influences; (b) toroidal magnetic field near the axis (blue dashed line) and line-integrated electron density (black solid line); (c) the magnetic fluctuation power spectra during 1.92-1.922s (red line) and 1.99-1.992s (black line).

In the experiment, it was also observed that the intensity variations of whistler waves at different harmonics are not consistent. As shown in Figure 6, (a) depicts the frequency spectrum during high-frequency electromagnetic fluctuations, (b) represents plasma density and toroidal field, and (c) displays the power spectrum of magnetic fluctuations during 1.92-1.922s and 1.99-1.992s. The intensity variations of whistler wave harmonics during 1.92s and 1.99s are distinctly different, with an increase in intensity for L2 and an opposite change in intensity for L1. This observation may be correlated with the plasma's absorption of whistler waves, but a detailed analysis awaits further investigation.

**Summary and discussion**

This article presents an experimental study of whistler wave instability driven by the temperature anisotropy of energetic electrons in steady-state ECW plasmas on a solenoid-free spherical torus. The study found that the frequency of the whistler wave is between 30-120 MHz, and is proportional to the Alfvén velocity. The whistler wave intensity is verified to be correlated with the electron density and $\Lambda$, with a density threshold for its excitation. In addition, the synergistic effect of LHW and ECW can suppress the whistler waves. The study also finds that plasma current disruptions occur in the spherical torus device with pure noninductive current drive when the whistler wave intensity exceeds a critical value and induces significant loss of energetic electrons. These findings suggest that ECW current-drive through the creation of energetic electrons may excite such waves, potentially limiting the efficiency of current drive. Finally, the interaction between whistler waves and low-frequency modes was observed, with the mode frequency positively correlated with the ion sound velocity. This could be a potential method for energetic particles to heat bulk ions through thermal ion Landau damping, and warrants further investigation.

The 100 kW ECW on EXL-50 can drive plasma currents of nearly 140 kA, often accompanied by whistler waves. Considering that RF also contributes to current drive, the study of whistler waves is of significant importance for understanding the physics of high-current drive in EXL-50 using ECW and for the application of high-current drive effects with ECW in other devices.

**Acknowledgments**

This work is supported by the High-End Talents Program of Hebei Province, Innovative Approaches towards Development of Carbon-Free Clean Fusion Energy (No. 2021HBQZYCSB006). The presented research is also supported in part by National Natural Science Foundation of China under Contract No. 12075284. We acknowledge the ENN team for supporting the experiments. All the data supporting the findings of this study are available from the corresponding author.